\documentclass[showpacs,superscriptaddress,amsmath,amssymb,aps,preprint]{revtex4}
\usepackage{graphicx}
\usepackage{bm}
\usepackage[utf8]{inputenc}
\usepackage{lmodern}
\begin{document}
\title{Quantum effects in beam-plasma instabilities} 
\author{A. Bret}
\affiliation{ETSI Industriales, Universidad de Castilla-La Mancha, 13071 Ciudad Real, Spain}
\affiliation{Instituto de Investigaciones Energeticas y Aplicaciones Industriales,
Campus Universitario de Ciudad Real, 13071 Ciudad Real, Spain}
\author{F. Haas}
\affiliation{Departamento de F{\'i}sica, Universidade Federal do Paran\'a, 81531-990, Curitiba, Paran\'a, Brazil}
\begin{abstract}
Among the numerous works on quantum effects that have been published in recent years, streaming instabilities in plasma have also been revisited. Both the fluid quantum and the kinetic Wigner-Maxwell models have been used to explore quantum effects on the Weibel, Filamentation and Two-Stream instabilities. While quantum effects usually tend to reduce the instabilities, they can also spur new unstable branches. A number of theoretical results will be reviewed together with the implications to one physical setting, namely the electron driven fast ignition scenario.
\end{abstract}
\maketitle
\section{Introduction}

\begin{figure}
\includegraphics[width=0.85\textwidth]{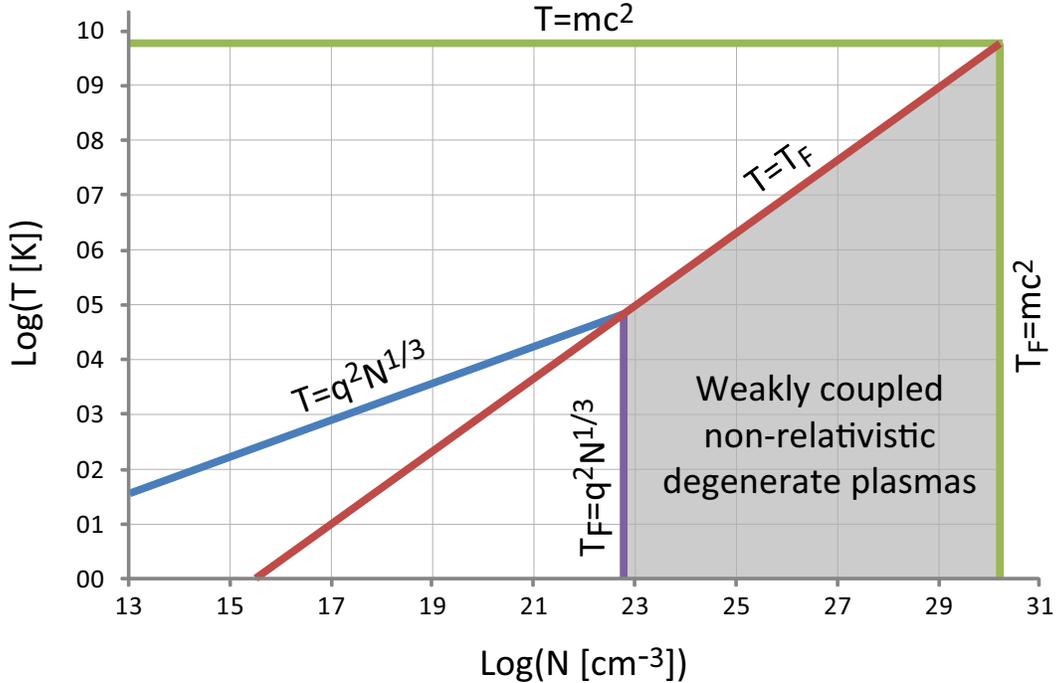}\\
\caption{Weakly coupled non-relativistic degenerate plasmas. (Color online)}
\label{fig1}
\end{figure}

A quantum plasma is a plasma where quantum effects cannot be neglected. Such is clearly the case for degenerate plasmas, where temperature is smaller than the Fermi energy. Considering for simplicity an electron gas of density $N$ and temperature $T$ on a fixed positive background, the Fermi energy $E_F$ varies like $N^{2/3}$. The line $T=E_F$ defines (see Fig. \ref{fig1}) the limit between classical and degenerate plasmas.  A classical plasma becomes relativistic for $T>mc^2$ where $m$ is the electron mass. A degenerate plasma becomes so for $T_F>mc^2$. In addition to this classification, a plasma is said weakly coupled when the typical Coulomb potential energy $q^2N^{1/3}$ is smaller than the mean kinetic energy. For classical plasma, this turns to $q^2N^{1/3}>T$ while for degenerate ones, the inequality reads $q^2N^{1/3}>T_F$ which just determines a threshold density beyond which the plasma is weakly coupled. Degenerate plasmas thus have this interesting feature that they are all the more weakly coupled, perfect gas-like, that they are dense. Relativistic effects will not be discussed here, nor spin effects. Note that these later can trigger quantum effects even for \emph{non}-degenerate plasmas with $T>T_F$ \cite{BrodinPRL2008}.

According to the present nomenclature,``quantum'' and ``degenerate'' plasmas are therefore not necessarily equivalent. A ``degenerate'' plasma simply has $T<T_F$. A ``quantum'' plasma needs quantum theory to correctly describe its behavior. Degenerate plasmas are quantum, but some non-degenerate plasmas may be quantum as well.

The study of quantum plasmas can eventually be traced back to the first days of solid state physics, as certain of its aspects can be approached forgetting about the lattice nature of the background ions \cite{mermin}. The next motivations for quantum plasma studies came from the physics of dense astrophysical objects \cite{Chabrier2002} and from Inertial Confinement Fusion (ICF). Regarding  the later, an Deuterium-tritium target in pre-ignition conditions should meet $T\sim 10^7$ K with $N\sim 10^{25}$, placing it at the border of the degeneracy frontier.

The investigation of streaming instabilities in quantum plasma arises naturally when considering the multi-stream plasma model \cite{Dawson,Haas2000} and has been directly needed recently to deal with the Fast Ignition Scenario for ICF \cite{Tabak,BretPPCF2009}. Besides these motivations, streaming instabilities are part of any plasma physics textbook, and it is natural to revisit them from the quantum point of view.

Quantum plasmas have been so far mostly investigated in the non-relativistic weakly coupled regime (see \cite{Hakim1978} for a relativistic formalism), although ultra-cold quantum plasmas usually refers to the low temperature, strongly coupled, degenerate regime. Quantum streaming instabilities have been dealt with so far exclusively in the weakly coupled non-relativistic regime (shaded region on Fig. \ref{fig1}), and the present Review will equally focus on this case. As is the case for the description of a plasma, the theory can deal with the fluid or the kinetic level. In the later case, quantum plasmas are studied from the so-called Wigner-Maxwell system of equations where the kinetic Wigner equation \cite{Wigner,ManfrediCourant} replaces the Vlasov equation. At the fluid level, quantum corrections to the fluid equations can be introduced \cite{Haas2000}. We will first treat the fluid and then the kinetic results on streaming instabilities in quantum plasmas.

An instability consisting in the temporal exponential growth of some physical quantity, will by design promptly reach the non-linear regime and then saturation. Among the works already achieved on nonlinear effects in quantum plasmas (see \cite{ShuklaEliasson2010} for a review), none are dealing with the nonlinear phase of streaming instabilities. The present article is therefore restricted to what is know on their \emph{linear} phase.

Although the Weibel and the filamentation instability differ (see \cite{BretReview2010} for a discussion), they are frequently considered interchangeably in literature. While filamentation instability stands for unstable modes propagating perpendicularly to the flow of a counter-streaming system, the Weibel instability refers to unstable modes triggered in a anisotropic plasma without any drift. The present review focuses on streaming instabilities, and the Weibel one will not be detailed. Let us just emphasize in this respect that both quantum fluids \cite{HaasLazar2008} and kinetic treatments \cite{HaasWeibel2008,Haas_Shukla_Eliasson_2008,tsintsadze2009} of the ``true'' Weibel instability are now available in the literature.

\section{Fluid treatment}
Like in the classical case, a number of plasmas instabilities such as two-stream, filamentation or Weibel are accessible to the fluid level \cite{Ichimaru,Fried1959,Basu}. The fluid treatment basically relies on some quantum corrections terms to the Euler fluid equation. Writing it for the electrons, for example, gives
\begin{equation}\label{eq:Euler}
   \frac{\partial \mathbf{v}}{\partial
   t}+(\mathbf{v}\cdot\nabla)\mathbf{v}=-\frac{q}{m}\left(\mathbf{E} + \frac{\mathbf{v}\times\mathbf{B}}{c}\right)
   +\frac{\hbar^2}{2m^2}\nabla\left(\frac{\nabla^2\sqrt{n}}{\sqrt{n}}\right)
   -\frac{n_0v_T^2}{n}G\nabla\left(\frac{n}{n_0}\right)^3,
\end{equation}
where $n,\mathbf{v}$ are the density and velocity fields respectively, $n_0$ the equilibrium density, $v_T$ the thermal velocity and $G$ a function of the chemical potential \cite{Eliasson2008}. Quantum corrections obviously consist in the last two terms of the r.h.s. The first of these is the so-called Bohm term accounting for undulatory quantum aspects. The second one clearly accounts for temperature effects and arises from the moments of the Wigner equation when considering a finite temperature Fermi-Dirac statistics \cite{Eliasson2008}. Formal linearization of these two terms can be found in Ref. \cite{BretPPCF2009}.

\subsection{Two-stream instability}
The dispersion equation for the quantum two-stream instability has been derived in Refs. \cite{Haas2000,Dubinov2001}. For two counter-streaming symmetric electron beams of density $n_0$ and velocity $v_0$ it reads,
\begin{equation}\label{eq:TS_fluid}
   \frac{1}{(\Omega-Z)^2-H^2Z^2/4}+  \frac{1}{(\Omega+Z)^2-H^2Z^2/4}=1,
\end{equation}
with $\Omega=\omega/\omega_p$, $Z=kv_0/\omega_p$, $\omega_p$ being the electronic plasma frequency of a single beam. The $H^2$ term at the denominator, sometime refers to as ``quantum recoil'' defines an extra, purely quantum unstable branch, located at higher $k$'s than the classical one, which has been detailed in \cite{Haas2000} (see Fig. \ref{fig2}). These secondary unstable modes have been related to negative energy waves \cite{HaasBret2010}.

Two-stream modes are electrostatic Langmuir waves having the electrons oscillate along the streaming direction. This is why no theory of these modes when accounting for a flow aligned magnetic field $\mathbf{B}_0$ is needed, since the correcting $\mathbf{v}\times \mathbf{B}_0$ force vanishes at all orders.

\subsection{Filamentation instability}
The two-stream instability pertains to unstable electrostatic modes propagating along the flow. The filamentation instability, which has to do with electromagnetic modes propagating \emph{perpendicularly} to the flow, has been discussed in \cite{BretPoPQuantum2007} for the un-magnetized case, and \cite{BretPoPQuantum2008} for the magnetized one. Considering again counter-streaming electron beams on a fixed neutralizing ion background, the classical filamentation instability is known to display a growth-rate saturating to a finite value at large $k$'s \cite{Godfrey1975}. Quantum effects has been found to set un upper limit to the unstable $k$ range. If the two beam have a density ratio much smaller than one, the largest unstable $k$ takes the simple form \cite{BretPoPQuantum2007},
\begin{equation}\label{eq:Fila_km_fluid1}
   k_m^2=\frac{2m\omega_b}{\hbar},
\end{equation}
where $\omega_b$ is here the plasma frequency of the thinner beam. In the classical case, the largest growth rate is reached from $k\sim \omega_p/c$, $\omega_p$ being the plasma frequency os the denser beam. Quantum effects thus become important when,
\begin{equation}\label{eq:Fila_km_fluid}
   k_m<\frac{\omega_p}{c} \Leftrightarrow\frac{\hbar\omega_p}{m c^2}>2\frac{\omega_b}{\omega_p},
\end{equation}
showing effects can be pronounced for very asymmetric systems with $\omega_b\ll\omega_p$.

\begin{figure}[!hb]
\centering{\includegraphics[width=8.0cm,height=6.0cm]{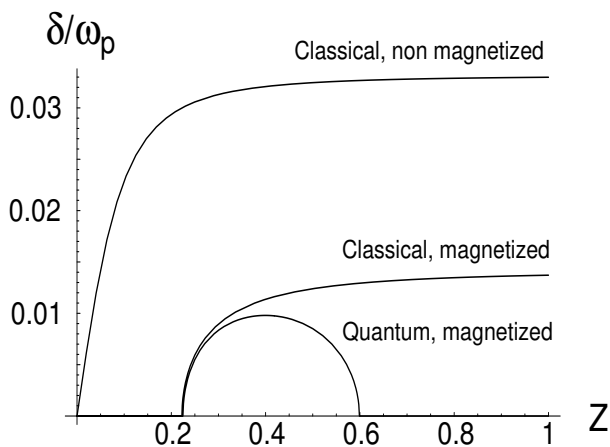}
\caption{Comparing the cold classical growth rate $\delta$ for the filamentation instability with the classical magnetized and quantum magnetized results. Parameters $v_0=0.1c$,  $\hbar\omega_p/mc^2=1.4~10^{-3}$ and $\Omega_B=3~10^{-2}$. The beams density ratio is 10.}
\label{fig:fila}}
\end{figure}

Filamentation instability has first been explored for the case of a flow aligned magnetic field in the cold approximation \cite{Godfrey1975}. The magnetized fluid theory accounting for the Bohm pressure term has been analyzed in Ref. \cite{BretPoPQuantum2008}. The guiding magnetic field is here measured through the dimensionless parameter,
\begin{equation}\label{eq:Omega_B}
   \Omega_B = \frac{q B_0}{m c\omega_p} \equiv \frac{\omega_B}{\omega_p}.
\end{equation}
Quantum effects have been found to emphasize the stabilization effects of the magnetic field. In the classical regime, magnetization introduces a low $k$ cutoff to the instability. In the quantum magnetized case, a large $k$ cutoff is added, as can be seen on Figure \ref{fig:fila}. As in the classical case, the instability can be completely suppressed beyond a given magnetic field which value is reduced by quantum effects.

\subsection{Full unstable spectrum}
\begin{figure}
\includegraphics[width=0.85\textwidth]{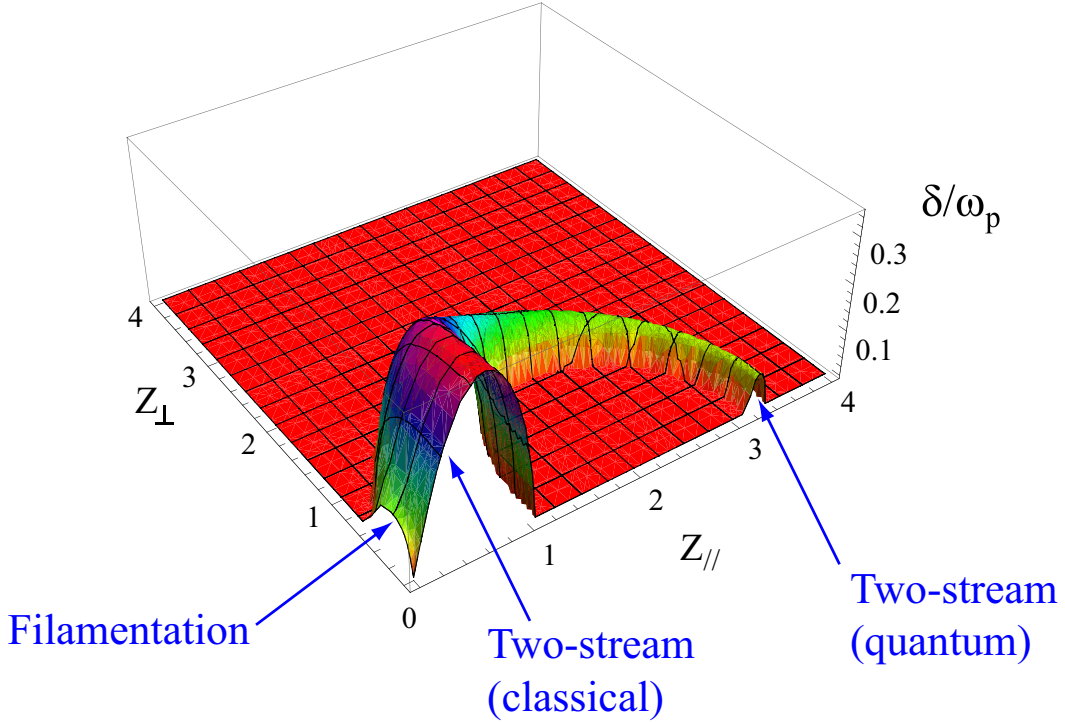}\\
\caption{Full unstable spectrum of a symmetric counter-streaming plasmas with velocities $v=0.6c$ with $\hbar\omega_p=0.6mc^2$. (Color online)}
\label{fig2}
\end{figure}

Investigating the full unstable spectrum for any kind of wave-vector orientations sheds an interesting light on the problem \cite{BretPRE2004,BretPRL2005,BretHaasPoP2010}. The basic idea is that unstable waves can be excited for intermediate orientations between the two-stream and filamentation propagation directions \cite{fainberg}. Linearizing the cold quantum fluid equations and computing the dielectric tensor for any kind of wave vectors allows for the derivation of the 2D unstable spectrum pictured on Fig. \ref{fig2} \cite{BretHaasPoP2010}. The growth rate of a given mode $\mathbf{k}$ has been systematically calculated in terms of $(Z_\parallel,Z_\perp)=(k_\parallel v_0/\omega_p,k_\perp v_0/\omega_p)$. Along the parallel axis, the classical two-stream unstable modes are easily recognized at low $Z$'s. At higher though, one now finds the purely quantum unstable branch. In the normal direction, the unstable range of filamentation modes is visible. Extending the calculation to the full spectrums uncovers a surprising connection between the two-stream branches. In the classical limit, the two-stream quantum branch is sent to $k_\parallel=\infty$, filamentation cutoff is sent to $k_\perp=\infty$, and the cold classical unstable domain fulfilling approximately $k_\parallel < 1$ is recovered.

\section{Kinetic treatment}
A kinetic approach is usually required to explore thermal effects on unstable modes. The basic reason for this is that this kind of modes are unstable in the cold limit because they can interact strongly with particles. If, thus, they are in phase with the particles in the cold limit, thermal effects will necessarily introduce a spread relevant for the wave-particle interaction. One can additionally infer from this physical picture that thermal effects usually reduce the instability, precisely because the thermal spread reduces the number of particles remaining in phase with the wave for a long time.

The expression of the quantum dielectric tensor for multi-species plasma reads (see \cite{BretHaasPoP2011} and references therein),
\begin{eqnarray}
\label{eps}
\varepsilon_{\alpha\beta} &=& \delta_{\alpha\beta}\,\left(1 - \sum_{j}\,\frac{\omega_{pj}^2}{\omega^2}\right)  \\ \nonumber  &+& \int\,d{\bf v}\,\frac{v_\alpha\,v_\beta}{\hbar\,(\omega - {\bf k}\cdot{\bf v})}\,\sum_{j}\,\frac{m_{j}\,\omega_{pj}^2}{\omega^2}\,\left(f_{j}^{0}\left[{\bf v} + \frac{\hbar\,{\bf k}}{2\,m_j}\right] - f_{j}^{0}\left[{\bf v} - \frac{\hbar\,{\bf k}}{2\,m_j}\right]\right),
\end{eqnarray}
where $m_j$, $\omega_{pj}$ and $f_{j}^{0}$ are the mass, plasma frequency and equilibrium distribution function for specie $j$ respectively. In principle, the expression above allows for the investigation of any linear phenomenon such as stopping power \cite{bret1993dielectric,bret1993stopping,bret1993straggling} or beam-plasma instabilities. Regarding the later, only the filamentation instability has been so far studied, for the case of two cold identical counter-streaming electrons beams on a fixed ions background \cite{BretHaasPoP2011}. Note that ``cold'' does not refer here to ``monokinetic'', but to the Fermi-Dirac distribution in the zero temperature limit.

\begin{figure}
\includegraphics[width=0.45\textwidth]{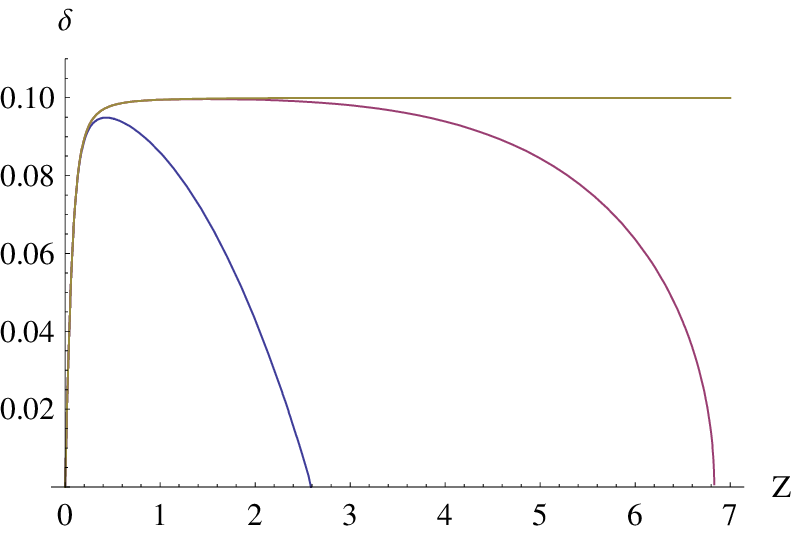}\includegraphics[width=0.45\textwidth]{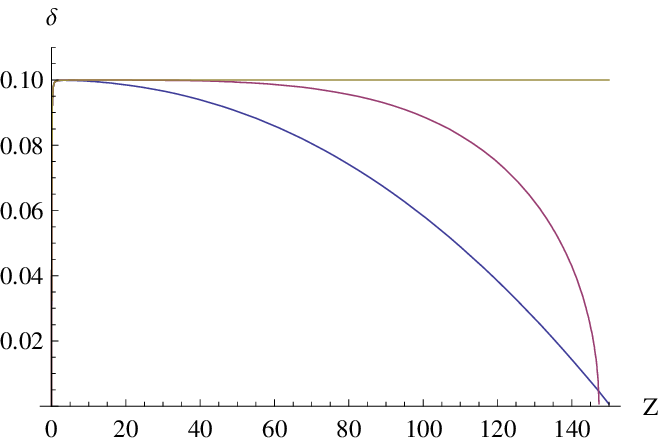}\\
\caption{Kinetic growth rate (blue) obtained solving the kinetic dispersion equation,
compared to the quantum fluid result with Bohm pressure term (purple) and to the classical cold
plasma result (yellow), in terms of the reduced wave vector $Z=kv_0/\omega_p$ for $v_0=0.1c$. Left: $v_0/v_F = 15$, and the fluid unstable range is wider than the kinetic one.
Right: $v_0/v_F = 900$, and the fluid unstable range is \emph{smaller} than the kinetic one. The saturation value
for the classical cold curve is simply $\beta=v_0/c$. (Color online)}
\label{fig4}
\end{figure}

For the case treated so far, the quadrature involved in the instability process can be calculated exactly, as the integrand is integrated over two Fermi spheres centered around $\pm \mathbf{v}_0$. Figures \ref{fig4} compare the growth rates in the cold classical, cold fluid quantum and cold kinetic quantum regimes. Both figures confirm the long wavelength equivalence of the 3 models. Noteworthily, while the Left plot gives, as expected, a fluid unstable range larger than the kinetic one, the situation is reversed on the Right plot.

It is interesting to compare the way some key quantities reach the classical limit in the quantum fluid and the quantum kinetic models. In both cases, the largest unstable wave vector tends to infinity like $\hbar^{-1}$. In the strong quantum limit $v_0/v_F \rightarrow 0$, both models equally yield a maximum growth rate $\delta_m$ approaching zero, but not the same way as the fluid one gives $\delta_m\propto (v_0/v_F)^{3/2}$ while the kinetic one yields $\delta_m\propto (v_0/v_F)^{2}$.

The general intuitive conclusion that the kinetic system should be more stable than the fluid one is thus recovered, in spite of the yet not understood ``anomaly'' observed on Fig. \ref{fig4} (right) where the kinetic unstable range is larger than the fluid one.

\section{Conclusion}
As stated in the introduction, quantum effects on streaming instabilities have been so far mainly studied from a fundamental point of view. However, an important practical motivation for their investigation in recent years has been the so called Fast Ignition Scenario (FIS) for Inertial Confinement Fusion (ICF). In the ``traditional'' scenario for ICF \cite{Lindl1995}, a $\sim 1$ mm pellet of Deuterium and Tritium is compressed \emph{and} heated by one single driver (mainly laser or heavy-ions). The heating requires an highly symmetrical compression to limit the growth of the Rayleigh-Taylor and Richtmyer-Meshkov instabilities \cite{PirizWouchuk,piriz1997,Huete2011}. The current solution, implemented at the National Ignition Facility, consists in converting the megajoule laser energy to an homogenous X-ray ``bath'' which compresses the target. In order to save the driver energy lost in the X-ray conversion, a scheme was proposed in 1994 decoupling the compression from the heating phase  \cite{Tabak}. While the compression phase is still achieved through a few hundreds kilojoule laser, the heating is performed  shooting a petawatt laser on the pre-compressed target. This later interaction produces a beam of relativistic electrons which, if properly tailored, deposit their energy at the center of the target a create the ``hot spot''. Since the target center is partially degenerate, the beam-plasma interaction at this location requires the inclusion of quantum effects. These later have been so far found negligible \cite{BretPPCF2009}, so that the energy lost exciting unstable modes can be evaluated near the target centered assuming a classical plasma.

Further applications should arise in astrophysics were dense objects are commonplace. Meanwhile, the theory still needs to extend. As we have seen, only the filamentation instability has been examined in the kinetic regime. Quantum relativistic extensions are equally possible \cite{MelroseQP,Mendonca2011,mendoncaPoP2011}.

{\bf Acknowledgments}\\
This work was supported by projects ENE2009-09276 of the Spanish Ministerio de Educaci\'{o}n y Ciencia and PEII11-0056-1890 of the Consejer\'{\i}a de Educaci\'{o}n y Ciencia de la Junta
de Comunidades de Castilla-La Mancha, and by CNPq (Conselho Nacional de Desenvolvimento Cient\'{\i}fico e Tecnol\'{o}gico).

\end{document}